\documentclass{raa}            % referee version: for submission

\usepackage{graphicx,times,epstopdf}             %for PS/EPS graphics inclusion, new
\usepackage{natbib}
\usepackage{amssymb,amsmath}
\bibpunct{(}{)}{;}{a}{}{,}
\usepackage{url}

\usepackage[a4paper=true]{}

\begin{document}

   \title{The Astrometric Performance Test of 80-cm Telescope at Yaoan Station and Precise CCD Positions of Apophis
%\footnotetext{$*$ Supported by the National Natural Science Foundation of China.}
}
%   \subtitle{I. Place Your Subtitle Here}

   \volnopage{Vol.0 (20xx) No.0, 000--000}      %%preserved for Editor. DOn't remove!
   \setcounter{page}{1}          %%starting page, preserved for Editor. DOn't remove!

   \author{Bifeng Guo
      \inst{1,2}
   \and Qingyu Peng
      \inst{1,2}
   \and Ying Chen
      \inst{1,2}
         \and Zhongjie Zheng
      \inst{1,2}
         \and Yijia Shang
      \inst{1,2}
          \and Dan Li
      \inst{1,2}
         \and Xiao Chen
      \inst{1,2}
   }

   \institute{Department of Computer Science, Jinan University, Guangzhou 510632, China; {\it tpengqy@jnu.edu.cn}\\
        \and
             Sino-French Joint Laboratory for Astrometry, Dynamics and Space Science, Jinan University, Guangzhou 510632, China\\
\vs\no
   {\small Received~~20xx month day; accepted~~20xx~~month day}}

\abstract{ The 80-cm azimuthal telescope is newly mounted at Yaoan Station, Purple Mountain Observatory in 2018. The astrometric performance of the telescope is tested in the following three aspects. (a) The geometric distortion of its CCD attached. It is stable in both a single epoch and multi epochs. Eight distortion solutions are derived over about one year. The maximum values range from 0.75 to 0.79 pixel and the median values range from 0.14 to 0.16 pixel. (b) The limit magnitude of stars. About 20.5 magnitude (Gaia-G) stars can be detected with Johnson-V filter exposured in 300 seconds. The astrometric error of about 20.5 magnitude stars is estimated at 0.14 arcsec using the fitted sigmoidal function. (c) The astrometric accuracy and the precision of stacked fast-moving faint object. 24 stacked frames of the potentially hazardous asteroid (PHA) (99942) Apophis are derived on April 14 and 15, 2021 (fainter than 18 mag) based on the ephemeris shifts. During data reduction, the newest Gaia EDR3 Catalog and Jet Propulsion Laboratory Horizons ephemeris are referenced as theoretical positions of stars and Apophis, respectively. Our results show that the mean (O-C)s (observed minus computed) of Apophis are -0.018 and 0.020 arcsec in right ascention and declination, and the dispersions are estimated at 0.094 and 0.085 arcsec, respectively, which show the consistency of the stacked results by \textit{Astrometrica}.
}

   \authorrunning{Bifeng Guo, Qingyu Peng, Ying Chen, et al }          %author_head in even pages
   \titlerunning{The Performance Test of 80-cm Telescope at Yaoan Station}  % title_head in odd pages

   \maketitle
%% The author head (on even pages) and the title head (on odd pages) will be
%% automatically extracted from \author{} and \title{}. Whenever the title is too long,
%% you will be asked to supply a shorter one by inserting either \authorrunning{} or
%% \titlerunning{} before \maketitle. Anyway, you can specify your own heads.
%%
%%
%% Note: In the following text body of your manuscript, please note several differences from
%%       other major journals:
%% (1) \subsection{Please Capitalize the First Letter of Each Notional Word in Subsection Title}
%% (2) Please Capitalize the First Letter of Each Notional Word in all tables' captions
%
%________________________________________________ sections below
%
\section{Introduction}           %% first-level sections will be auto-capitalized
\label{sect:intro}

  The 80-cm azimuthal-mounting telescope at Yaoan Station, Purple Mountain Observatory has served for precise astrometry since it was set up in the year of 2018. Although some precise positions by the telescope have been published (\citealt{2021A&A...645A..48Y}, \citealt{2021SSPMA..51b9502L}), the performance of the telescope has not been specified. The potentiality of the telescope should be taped through the performance test, which may serve for some survey or occultation observation. In this paper, we aim to test its astrometric performance in the following three aspects. (a) The geometric distortion of its CCD attached. (b) The limit magnitude of stars. (c) The astrometric accuracy and the precision of stacked fast-moving faint object.

  The astrometric potentiality of the telescope can be tapped by the solutions for the geometric distortions (GD). The \textit{Hubble Space Telescope} (\textit{HST}) is an obvious example, whose \textit{WFPC2} chip has a maximum geometric distortion of about 5 pixels at the edge of its field (\citealt{2003PASP..115..113A}). By increasing the accuracy of the linear terms, a more accurate solution for GD is determined. After the GD correction, the observations of Saturnian satellites have much better precision than ever before (\citealt{2006PASP..118..246F}). Later, the \textit{HST} observations of M92 are taken as a distortion-free reference frame to improve the GD solution of Keck II 10-m telescope's near-infrared camera (NIRC2)
  in its narrow field mode, which is a major limitation for the proper motion measurements of Galactic central stellar cluster (\citealt{2010ApJ...725..331Y}).

   The limit magnitude of the telescope is also the embodiment of the astrometric potentiality. Namely, it tells us how faint the telescope can observe the object.   Usually, it guides us to choose the objects of interest and develop observational plans. For example, we can explore the physical properties of the interested faint stars with high-precision astrometric positions. Also, some interesting faint Kuiper-belt objects can be detected and explored.

   We explore the potentiality of observing fast-moving faint object by analyzing the  astrometric accuracy and precision of stacked observations of Apophis. Besides, the astrometric positions of Apophis are also precious. Apophis
   was discovered by Bernardi, Tholen and Tucker on June 19, 2004 at Kitt Peak observatory (Minor Planet Supplement 109613). In the same year, the impact probability reached 2.7$\%$ in April 13, 2029 encountering with Earth, but later it was ruled out (\citealt{2008Icar..193....1G}, \citealt{2013Icar..224..192F}). Nevertheless, some other close encounter events are forecast in 2068, 2085 and 2088 (\citealt{2018A&A...617A..74S}). Successive observations remain needed to monitor the positions of Apophis, which is of great importance to improve its orbital parameters. Besides, the observations in 2021 are the key information of analyzing the Yarkovsky effect for such a small-mass object (\citealt{2015Icar..252..277V}, \citealt{2018Icar..300..115B}). According to Jet Propulsion Laboratory Horizons ephemeris\footnote{\it http://ssd.jpl.nasa.gov/} (JPL), although the predicted visual magnitude of Apophis is usually fainter than 20 mag in recent years, it is brighter than 17 mag in February, 2013 when it is a valuable oppotunity to observation. Some researchers (\citealt{2015MNRAS.454.3805W}, \citealt{2015A&A...583A..59T}) seized the oppotunity and obtained the valuable precise positions of Apophis.

Because of its fast motion and faint brightness, high-quality observations of Apophis are hard to obtain with a small-aperture ground-based telescope. Observing in short-time exposures and stacking the frames are a good solution to obtain high-quality images, by which the performance of the telescope can be tested. This method was first proposed by \cite{1992AAS...181.0610T} to detect faint Kuiper-belt objects a few decades ago. Later, this technique is well used to survey and find faint near-Earth asteroids (NEAs) (\citealt{2014ApJ...782....1S}, \citealt{2014ApJ...792...60Z}).  It is showed that the technique can discover asteroids 10 times fainter than conventional searches (\citealt{2015AJ....150..125H}). \citealt{2017AcASn..58...49W} use the iterative stacking method to detect faint asteroids. Subsequently, \citealt{2020AcASn..61...28L} finds that shift-and-add method can also improve the mesurement of astrometric positions for some faint satellites of Jupiter. The developed open software also provides the function of stacking like \textit{Astrometrica}\footnote{\it http://www.astrometrica.at/} and \textit{MaxIm DL}\footnote{\it https://diffractionlimited.com/maxim-dl/}. This paper explores an alternative stacking way based on the ephemeris shifts of Apophis to obtain the precise positions of it.

The contents of this paper are arranged as follows. In Section \ref{sect:Telescope Performance}, we give the specification and the operation performance of the telescope and its CCD attached. In Section \ref{sect:Astrometric Test of the telescope}, we elaborate the performance test of the telescope in geometric distortion, limit magnitude and the stack of fast-moving object Apophis. The conclusions and outlooks are shown in the last section.
%% Authors can give a citation as 'Michel et al. 1992'.
%% You may also use \cite, \citep and \citet for citation, and use Table~1 or Figure~1
%% and so forth. Using \ref and \label for cross-references of Tables/Figures
%% is a good way in adjusting/adding/removing text, tables or figures.

\section{Introduction of the Telescope}
\label{sect:Telescope Performance}

\subsection{Specifications of the Telescope and CCD}

The 80-cm azimuthal-mounting telescope at Yaoan Station, Purple Mountain Observatory is located at E101${\degr}$ 10${\arcmin}$ 51.0${\arcsec}$, N25${\degr}$ 31${\arcmin}$ 43.0${\arcsec}$, whose IAU code is O49. The specifications of telescope and the CCD attached are shown in Table \ref{Tab:tab1}.

%% Tab1
%
%               one-column-spanning table
%________________________________________ Table 2: Use_of_the routines

\subsection{Operation Performance}

 The azimuthal-mounting telescope is well-known to easily construct than the equatorial-mounting telescope and is more stable for a large telescope. To follow the rotation of the sky, both of the axes (azimuth axis and altitude axis) must be turned around with a changing angle speed, which is compensated for photography. However, if the pointing of the telescope goes close to the zenith, its azimuth will change 180${\degr}$ in a short time and there is a small region where the observations are impossible.

     \begin{table}
	\begin{center}
		\caption[]{ Specifications of the Telescope and its  CCD}\label{Tab:tab1}
		
		%%Please Capitalize the First Letter of Each Notional Word in table's caption
		
		\begin{tabular}{cccc}
			\hline\noalign{\smallskip}
			Items &  Parameters                  \\
			\hline\noalign{\smallskip}
			Focal Length  & 800cm               \\
			Diameter of Primary Mirror  & 80cm  \\
			F Ratio  &         10            \\
			CCD Field of View  & 11.8$'$ $\times$ 11.8$'$       \\
			Size of Pixel  & 13.5$\mu$m $\times$ 13.5$\mu$m \\
			Size of CCD Array  & 2048 $\times$ 2048     \\
			Angular Resolution  & 0.346$''$/px                 \\
			Pointing Accuracy  & $ < $8$''$ RMS                \\
			Tracking Speed  & 13${\degr}$ per second                \\
			Tracking Accuracy  & $ < $0.25$''$ in 5 minutes                \\
			\noalign{\smallskip}\hline
		\end{tabular}
	\end{center}
\end{table}

 \section{Astrometric Test of the telescope}
 \label{sect:Astrometric Test of the telescope}

 To explore the astrometric potentiality of the telescope, we test the operation performance of the azimuthal-mounting telescope from the obtained observations in geometric distortion (GD), limit magnitudes and the stack of fast-moving object Apophis. Few stars may be observed in the sparse sky areas, which leads to the worse accuracies and precisions due to the constructed plate model (\citealt{1980A&A....89...41L}). Generally, the calibration field is observed before observing the field of sparse sky areas so that the geometric distortion can be solved. In this work, we focus on the crowded sky areas.

 \subsection{The Stability of Geometric Distortion}

 Two kinds of observations are carried out to test the stability of GD. One is for single epoch observation data. In this case, multiple images are taken on the same night with the stars always at the same pixel location. The highest possible astrometric precision and the stability of GD during one night can be tested. Another is for multi-epoch observation data. These observations allow us to test the GD variation over time (nights, months, weeks and years).

\begin{figure}
	\centering
	\includegraphics[width=.9\textwidth, angle=0]{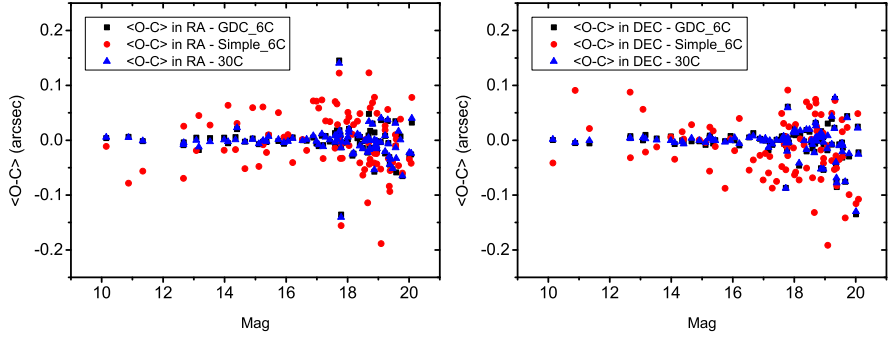}
	\caption{The left and right panel show the 3 different reduced method of the accuracies of the stars in right ascention and declination, respectively.}
	\label{GDC_OMC}
\end{figure}

\begin{figure}
	\centering
	\includegraphics[width=.9\textwidth]{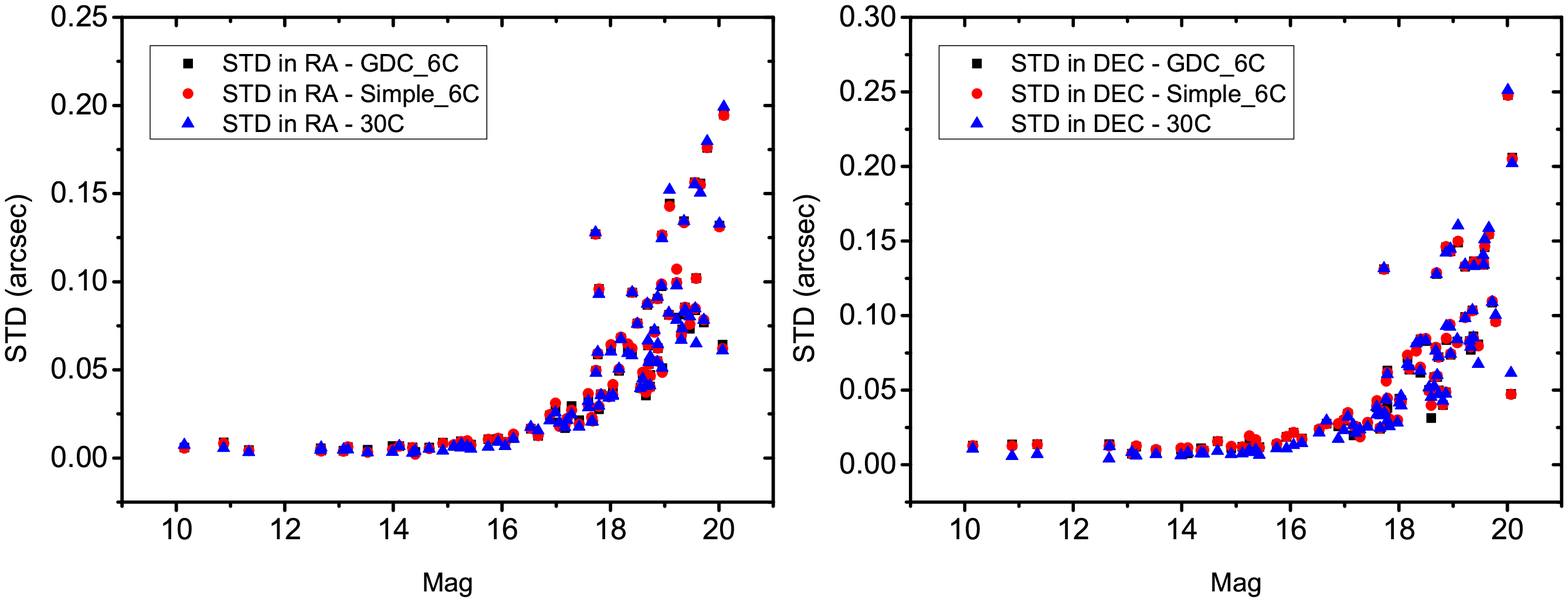}
	\caption{The left and right panel show the 3 different reduced method of the precisions of the stars in right ascention and declination, respectively.}
	\label{GDC_STD}
\end{figure}

 For the stability of GD in the same epoch (the same night), we first solve the GD of CCD by the observations of open cluster M35 with Johnson-I filter on Nov 28, 2019 using a convenient method (\citealt{2012AJ....144..170P}). Then we retrieve the observations taken at the same night with the stars always at almost the same pixel location (some pixels' shifts exist due to the tracking error of the telescope). The pointing of the telescope is at (05$ ^{h} $ 41$ ^{m} $ 43.0$ ^{s} $, -01${\degr}$ 50${\arcmin}$ 30.0$''$, J2000) with exposure time 60 seconds for each CCD frame and totally 20 frames are taken. Three methods are used to perform data reduction for the stars to investigate the stability of GD. (a) A simple 6-parameter model transformation. (b) A 6-parameter model transformation after GD correction. (c) A 30-parameter model transformation. Fig. \ref{GDC_OMC} and Fig. \ref{GDC_STD} show the accuracies (mean (O-C)s) and precisions (STDs) of the reduced results, respectively. From the results, the mean (O-C)s using 6-parameter model after GD correction improve significantly, which is similar to the 30-parameter model. And the three methods above have almost the same precisions. The results show that the geometric distortion is stable during this epoch.

\begin{figure}
	\centering
	\includegraphics[width=.9\textwidth, angle=0]{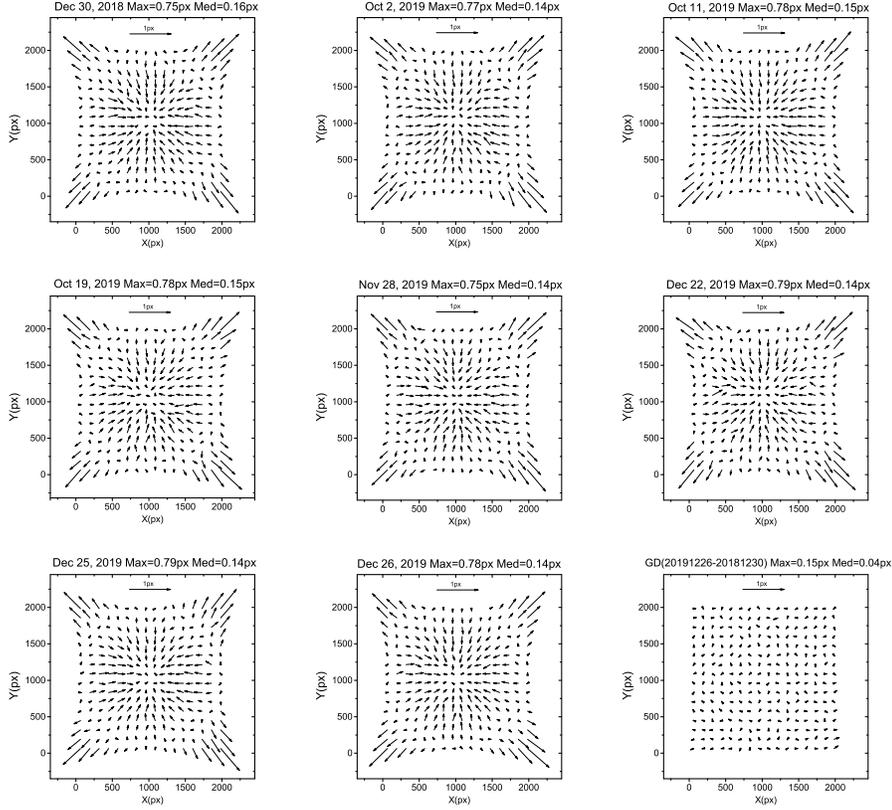}
	\caption{The GD graphs are derived from the observations of open cluster M35 or NGC2324. A factor of 500 is used to exaggerate the magnitude of each geometric distortion vector. The first eight panels show the GD graphs in different epoch. The last panel shows the difference of the two geometric distortion solutions of Dec 26, 2019 and Dec 30, 2018. }
	\label{GD_graphs}
\end{figure}

 For the stability in different epochs (different nights), it's an effective way to test with the stars locating on the same pixel location. After GD correction, other systematics effects of the stars (assuming zero proper motion) from multi-epoch data sets can be estimated. For the time arrangement of the telescope, such proper observations haven't been taken. In our work, we solve the distortion for eight nights (range from Dec 30, 2018 to Dec 26, 2019) including the time of adjacent nights, weeks, months and years to test the stability of the distortion (see Fig. \ref{GD_graphs}). In Fig. \ref{GD_graphs}, the first eight panels show the GD vector graphs of the CCD solved by the observations of the open cluster M35 or NGC2324. They are all taken with Johnson-I filter. For the details of the GD vector graphs, the large values of GD locate at the four corners of the CCD, which point to the corresponding corner, while the intermediate values mainly locate around the center of the CCD, which point to the CCD center. There is an annular region of the CCD where the GD values are small and the pointings are random. From the first eight panels, the shapes of the vector graphs show the consistency and the maximum values of the distortions range from 0.75 to 0.79 pixel over about one year. The median values of the distortion range from 0.14 to 0.16 pixel. Our results show that the distortion is stable over time. While the last panel of Fig.\ref{GD_graphs} shows the difference of the two geometric distortion solutions of Dec 26, 2019 and Dec 30, 2018. The difference of GD values of each region in the pixel coordinate are stable and the median value is 0.04 pixel (about 0.014 arcsec). For the stable GD solutions, it is easier to obtain precise positions of the objects after GD correction even if there aren't enough stars in the field of view to construct high-order plate model.

\begin{figure}
	\centering
	\includegraphics[width=.9\textwidth, angle=0]{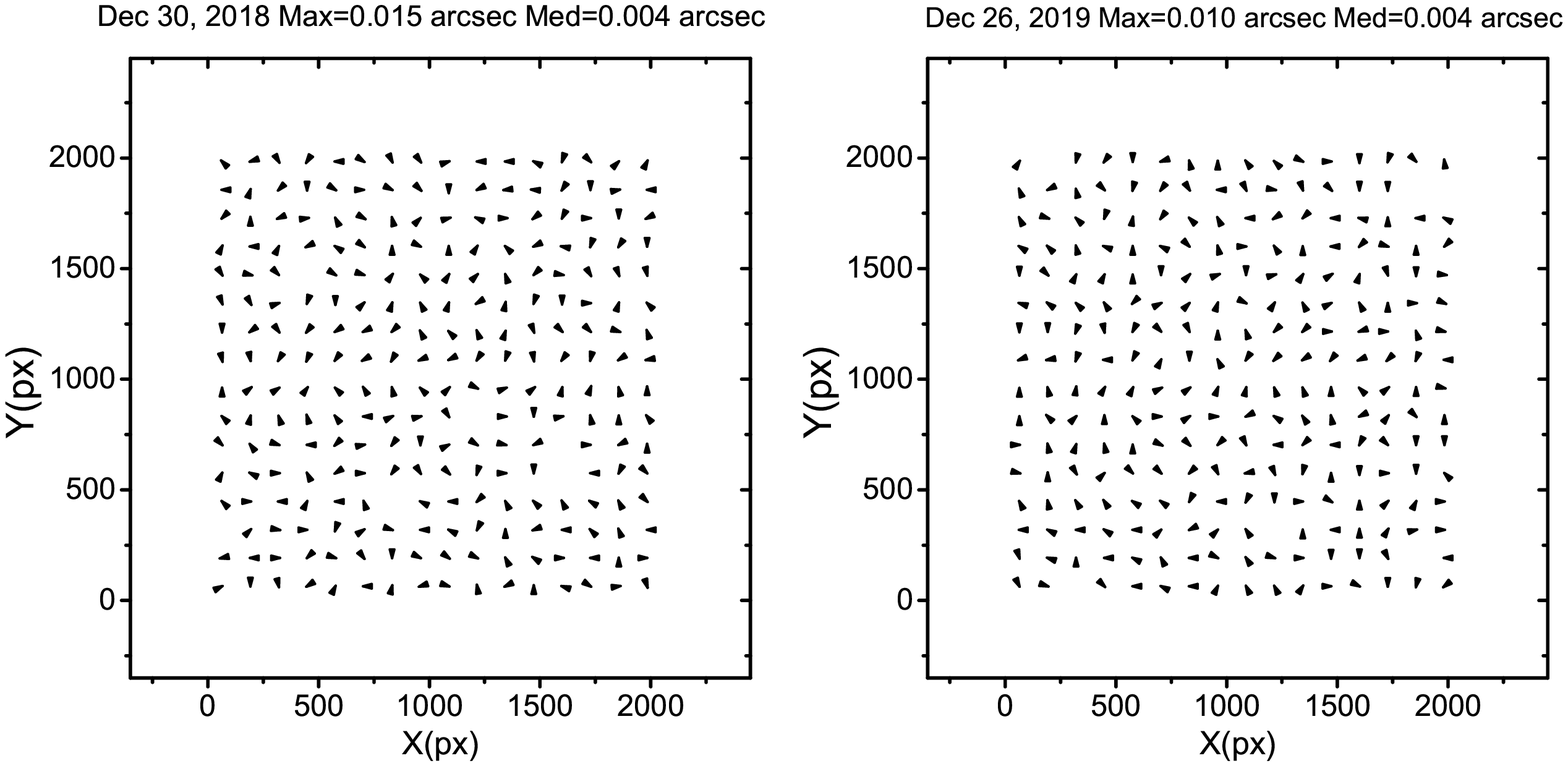}
	\caption{The left and right panels show the residual graphs derived from the observations of open cluster M35 on Dec 30, 2018 and Dec 26, 2019, respectively .A factor of 1,000 is used to exaggerate the magnitude of each vector.}
	\label{OMC_res}
\end{figure}

 We have also derived the distributions of the astrometric errors of the calibration field for open cluster M35, which are also used to derive the geometric distortion solutions. We perform data reduction with 30-parameter model and compute the mean residual (O-C)s of each area in right ascention (X-axis) and declination (Y-axis), which are shown in Fig.\ref{OMC_res}. The resultant values ($\sqrt{\sigma_{ra}^2+\sigma_{dec}^2}$) of the residuals are very small and the obvious trend of the distribution hasn't been found. The median value of the residuals is 0.004 arcsec.

\begin{table}
	\begin{center}
		\caption[]{The specifications of the two sets of observations to explore the limit magnitude of the telescope. The first two columns list the observation dates and the obtained number of frames of stars. The last four columns shows the paramters of the fitted sigmoidal functions for the astrometric precisions.}\label{Tab:tab2}
		
		%%Please Capitalize the First Letter of Each Notional Word in table's caption
		
		\begin{tabular}{cccccccccc}
			\hline\noalign{\smallskip}
			Date & Frames & Filter & Exptime(s) & Limit Mag & Para-A1 & Para-A2 & Para-X0 & Para-dx        \\
			\hline\noalign{\smallskip}
			Nov 5, 2018 & 8 & Johnson V & 300 & ~20.5 & 0.00748   & 0.15577  & 19.30018 & 0.59252      \\
			Oct 9, 2019 & 4 & Johnson V & 300 & ~19.0  & 0.01032   & 0.18495  & 18.56849 & 0.75234     \\
			
			\noalign{\smallskip}\hline
		\end{tabular}
	\end{center}
\end{table}

  \subsection{The Limit Magnitude}

\begin{figure}
 	\centering
 	\includegraphics[width=.9\textwidth, angle=0]{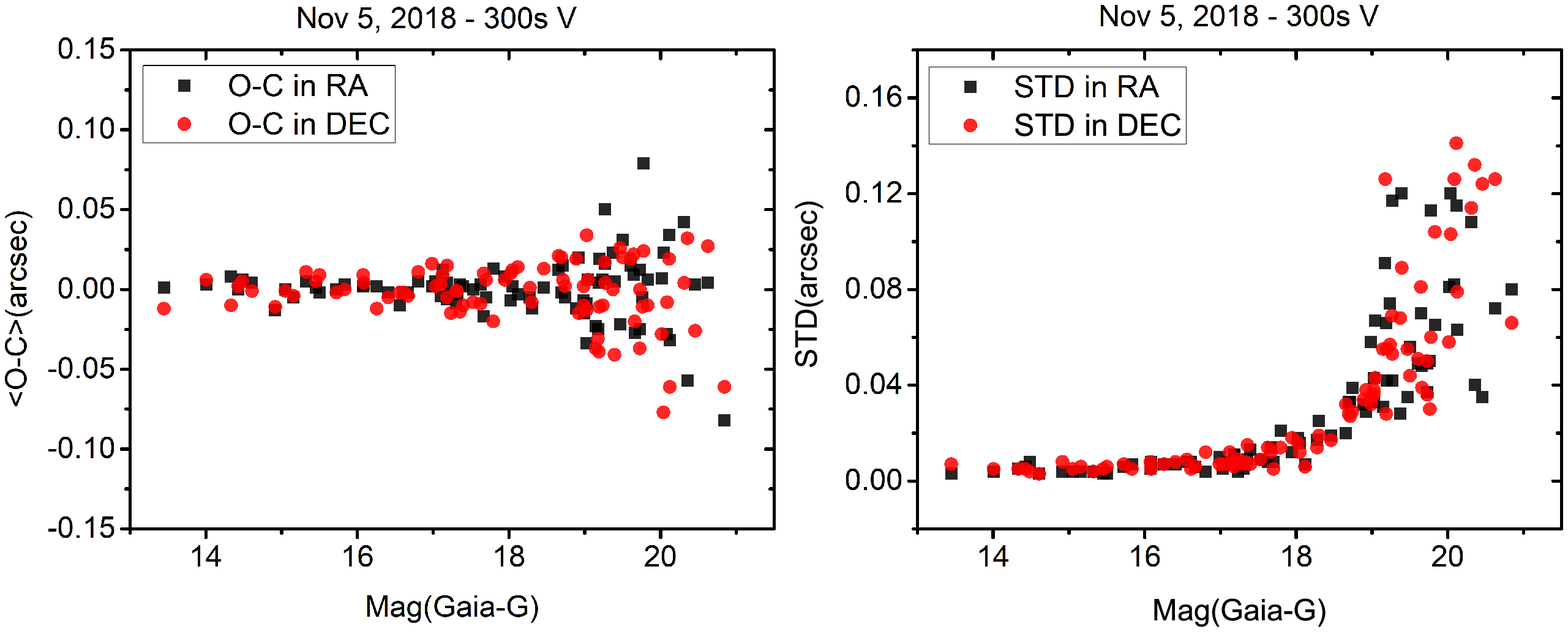}
 	\caption{The left and right panel show the accuracies and the precisions of the stars in right ascention and declination on November 5, 2018, respectively.}
 	\label{Fig1}
 \end{figure}

\begin{figure}
 	\centering
 	\includegraphics[width=.9\textwidth, angle=0]{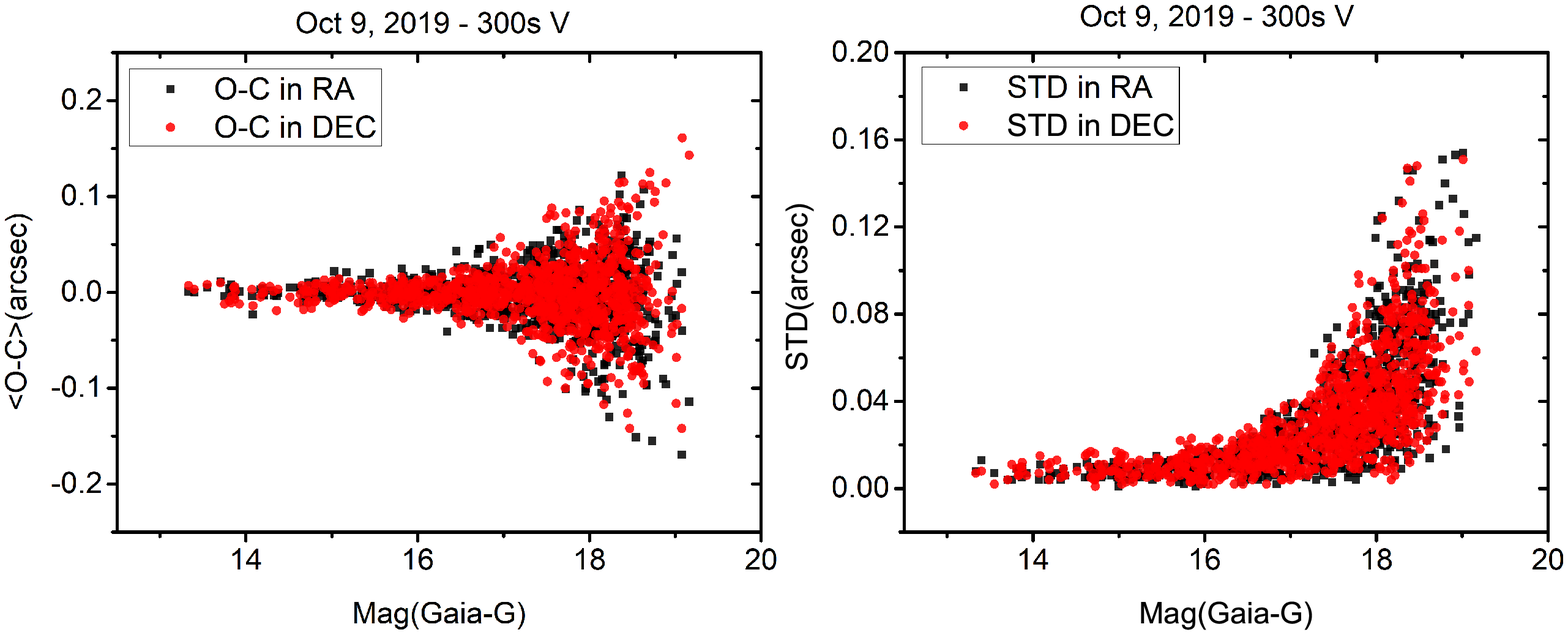}
 	\caption{The left and right panel show the accuracies and the precisions of the stars in right ascention and declination on October 9, 2019, respectively.}
 	\label{Fig2}
 \end{figure}

 The limit magnitude of the stars is also explored by reducing the observations (reduction details in Section 3.3.6) with long exposure time (300 seconds). The relationship between astrometric errors and exposure time can be consulted from Lindergren's work (\citealt{1980A&A....89...41L}). The specifications of the used observations are shown in Table \ref{Tab:tab2}. Fig.\ref{Fig1} and Fig.\ref{Fig2} show the reduction results of the observations and the pointings of the telescope are at (23$ ^{h} $ 01$ ^{m} $ 15.79$ ^{s} $, -07${\degr}$ 20${\arcmin}$ 17.56$''$, J2000) and (19$ ^{h} $  02$^{m} $ 24.72$ ^{s} $, -22${\degr}$ 51${\arcmin}$ 03.88$''$, J2000), respectively. From the results over two nights, the stars fainter than 20 magnitude (Gaia-G) can be detected using gaussian centering algorithm where the detection threshold is set as 3 $ \sigma $ of the background in each frame. The signal-noise ratio (SNR) of the faintest stars of each frame is about 5.

 \begin{figure}
 	\centering
 	\includegraphics[width=.9\textwidth, angle=0]{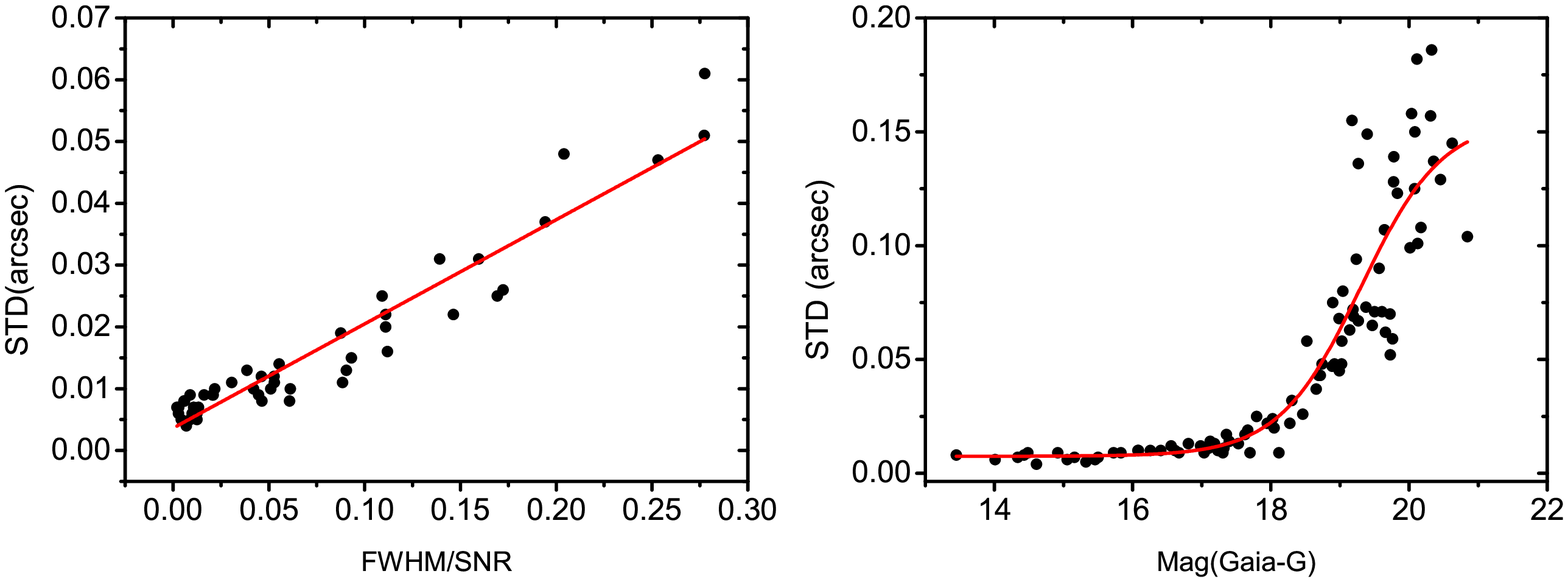}
 	\caption{The left and the right panel show that the astrometric errors fit with FWHM/SNR (SNR$>$10) using linear function and Gaia-G magnitude using sigmoidal function, respectively. The fitted functions are shown in red line.}
 	\label{mag_fit}
 \end{figure}

 The relationship between astrometric errors and FWHM/SNR is really an effective way to estimate the astrometric precision, especially for high-SNR stars. Considering both bright and faint stars across the field of view, it is found that the sigmoidal function can fit well with Gaia-G Magnitude and the astrometric precision (\citealt{2019MNRAS.490.4382L}). The expression of sigmoidal function is shown in formula \ref{eq:sigmoidal}. We have compared the two methods above to estimate the astrometric errors ($\sqrt{\sigma_{ra}^2+\sigma_{dec}^2}$). Using the observations on Nov 5, 2018, the astrometric errors can be fitted in Fig. \ref{mag_fit}. The left and the right panel show that the astrometric errors fit with FWHM/SNR (SNR$>$10) using linear function and Gaia-G magnitude using sigmoidal function, respectively. Both of them can be fitted well. Although the SNRs for faint stars can't be determined accurately, the precisions can be well expressed with a sigmoidal function. Therefore, we use the sigmoidal function to estimate the precision of the stars.

 For the observations in 2018, about 20.5 magnitude stars can be detected and the corresponding astrometric error is estimated at 0.14 arcsec using the fitted sigmoidal function. While for the observations in 2019, about 19.0 magnitude stars can be detected, the corresponding astrometric error is estimated at 0.12 arcsec. The difference of the limit magnitude over two sets of observations are mainly caused by the atmosphere conditions. The astrometric precisions of these observations are estimated with sigmoidal functions, whose parameters are shown in Table \ref{Tab:tab2}. However, brighter stars might have larger astrometric errors from the observations because they might be saturated. The sigmoidal function only takes the unsaturated stars into account.

\begin{equation}
	y = \dfrac{A1-A2}{1+e^{(X-X0)/dx}}+A2
	\label{eq:sigmoidal}
\end{equation}

\subsection{The Stack of Fast-Moving Object}

\label{sect:Alignment and Stacking}
In order to test the astrometric performance of the fast-moving object, we explore the astrometric accuracy and precision of the stacked observations of Apophis.

\subsubsection{Observations}

During the observation, the angular rate of Apophis is about 1 arcsec per minute (details in Fig. \ref{Fig5}). Given that the Full Width at Half Maximum (FWHM) of the star images are usually 3-5 pixels and the astrometric precision, a CCD frame of Apophis is required to be exposured within 20 s (about $ \frac{1}{5} $ $ \sim $ $ \frac{1}{3} $ FWHM). Due to the object is so faint and invisible in a frame, stacking method has to be performed to improve the signal-noise ratio. The observations overview of Apophis are shown in Table \ref{Tab:tab3}. The observations are taken on April 14 and 15, 2021 using Johnson I filter. The seeing over two nights are similar and the predicted visual magnitude of Apophis is fainter than 18 mag.

We evaluate the star images' FWHMs and the shifts in pixel with respect to its nearby CCD frame for each frame. We also derive the middle time of observation for each frame from the headers of FITS files. Given the stacked SNR of the object and the number of the total frames, we stack in group of 10 frames. To ensure the quality of stacked frames, we select frames with similar FWHMs of star images, near observation time and small pixel shifts into a group. Finally, 13 and 11 groups of observations were selected on April 14 and 15, 2021, respectively.

%% Table 2
\begin{table}
	\begin{center}
		\caption[]{Observations Overview. The first two columns list the observation dates and the obtained number of frames of Apophis. The 5th column shows the stars' FWHMs ranges evaluated from each frame and the last column shows the predicted visual magnitude of Apophis from JPL ephemeris.}\label{Tab:tab3}
		
		%%Please Capitalize the First Letter of Each Notional Word in table's caption
		
		\begin{tabular}{cccccccc}
			\hline\noalign{\smallskip}
			Date & Frames & Filter & Exptime(s) & FWHM(px) & Predicted Mag       \\
			\hline\noalign{\smallskip}
			April 14, 2021 & 141 & Johnson I & 20 & 2.7-4.3 & 18.16           \\
			April 15, 2021 & 135 & Johnson I & 20 & 2.8-4.5 & 18.20           \\
			
			\noalign{\smallskip}\hline
		\end{tabular}
	\end{center}
\end{table}

%% Figure3-5 and Table 3
\begin{figure}
	\centering
	\includegraphics[width=.9\textwidth]{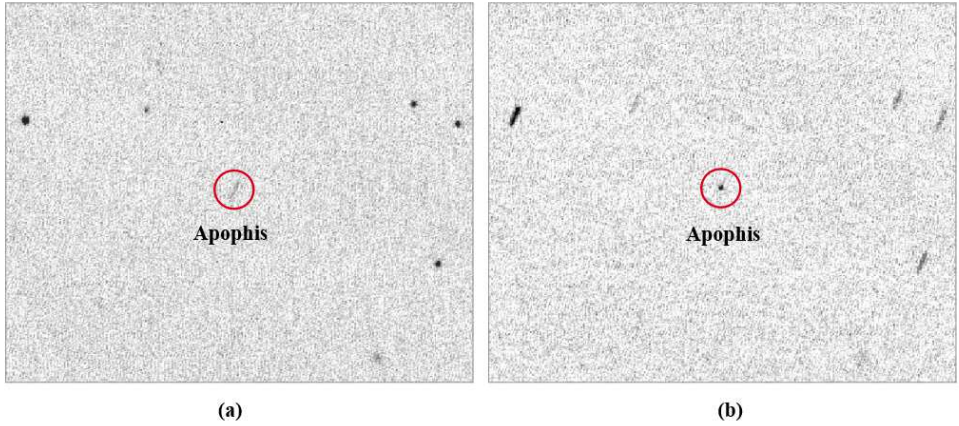}
	\caption{The left and the right panel are the subframes after stacking. The left shows a star-stacked subframe while the right shows an object-stacked subframe. }
	\label{Fig3}
\end{figure}

\begin{figure}
	\centering
	\includegraphics[width=.9\textwidth, angle=0]{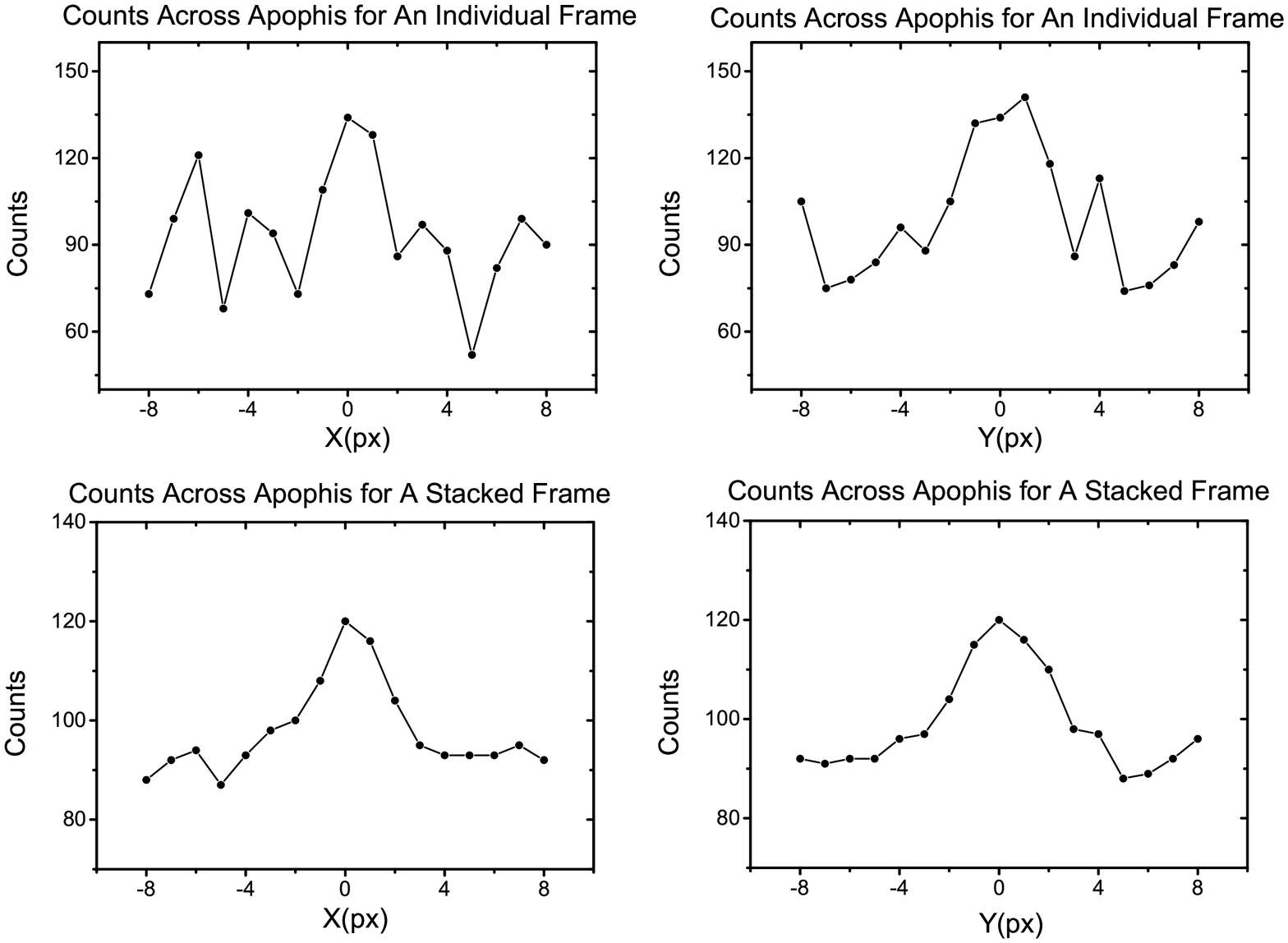}
	\caption{The top two panels show the counts (ADU) across pixel coordinate (in both X and Y axis) of Apophis in an individual frame. The bottom two panels show the counts (ADU) across pixel coordinate (in both X and Y axis) of Apophis in an individual frame. For each panel, X axis shows the pixel location of Apophis, while Y axis shows the pixel counts, where we define the center pixel location of Apophis is 0.}
	\label{Fig4}
\end{figure}

\subsubsection{Alignment of Stars}

In a group, we select a high-SNR frame as a master frame such that other frames in this group should be aligned to. A stacked frame refers to the frame after stacking, which is equivalent to a master frame taken by a larger-aperture telescope with the same exposure time. According to what we align to (the stars or the object), the stacked frames includes star-stacked frames and object-stacked frames. Flat fielding and bias substracting are made to remove the difference of pixel response to photoelectrons before stacking. The procedures of aligning stars are described as follows. Firstly, we obtained the position of each star image and performed photometry. Thus, the pixel positions $ (x, y) $ and the instrumental magnitude of each star image of the master frame can be obtained. In the same way, we can derive the positions $ (x${\arcmin}$,  y${\arcmin}$) $ and the instrumental magnitude of each individual frame. The same star locates in both master frame and individual frame in a group can be indentified conveniently through the relationships of the positions and the instrumental magnitudes. As for each individual frame, we can calculate the corresponding position of the master frame through a 6-parameter model transformation (see equation \ref{eq:1}) fitted by a least square method. We have tried higher order polynomial models (including 12-parameter model and 20-parameter model) to align the images for the observations of Apophis. However, no obvious improvement is found for the accuracy and the precision in this set of observations, which might reflect the stable atmosphere conditions and the geometric distortion. We also think that high-order polynomial model alignment will work better in the group of images with different pointings (dithered frames), different epochs or in a rapidly changing atmosphere.
 According to the transformed model, the pixel positions of each individual frame in a group can be transformed to the master frame using Drizzle method (\citealt{2002PASP..114..144F}).

\begin{equation}
 \left\{
 \begin{aligned}
	x & =ax' + by' + c \\
	y & =dx' + ey' + f \\
 \end{aligned}
\right. \label{eq:1}
\end{equation}

\subsubsection{Alignment of Object}

After each individual frame has been aligned to the master frame according to the positions of the stars, we are to calculate the shifts in pixel of each individual frame with respect to the master frame to align the object such as Apophis. Specifically, we first derive the middle of exposure time of both master frame and individual frames. Then, we can calculate the object's right ascention and declination (apparent positions) according to its ephemeris. In these frames (including the master frame and each individual frame), we can obtain the pixel positions $ (x, y) $ and the standard coordinate $($$\xi$,$\eta$$)$ of each star (more details in Section 3.3.6). The standard coordinate positions $($$\xi_{o}$,$\eta_{o}$$)$ of the object can also be calculated through the central prejection after considering the atmosphere refraction. A least squares scheme is used to solve the plate model with a 6-parameter model (see equation \ref{eq:2}). Next, we can derive the positions of the object in the pixel coordinate $ (x_{o}, y_{o}) $ of both master frame and each individual frame according to the plate model. Finally, the shifts in pixel of each individual frame with respect to the master frame $\Delta$$ x_{o} $ and $\Delta$$ y_{o} $ can be calculated and added to the parameter \textit{c} and \textit{f} in equation \ref{eq:1}, respectively after the alignment of stars.

\begin{equation}
	\left\{
	\begin{aligned}
		\xi & =ax + by + c \\
		\eta & =dx + ey + f \\
	\end{aligned}
	\right. \label{eq:2}
\end{equation}

\begin{figure}
	\centering
	\includegraphics[width=.9\textwidth, angle=0]{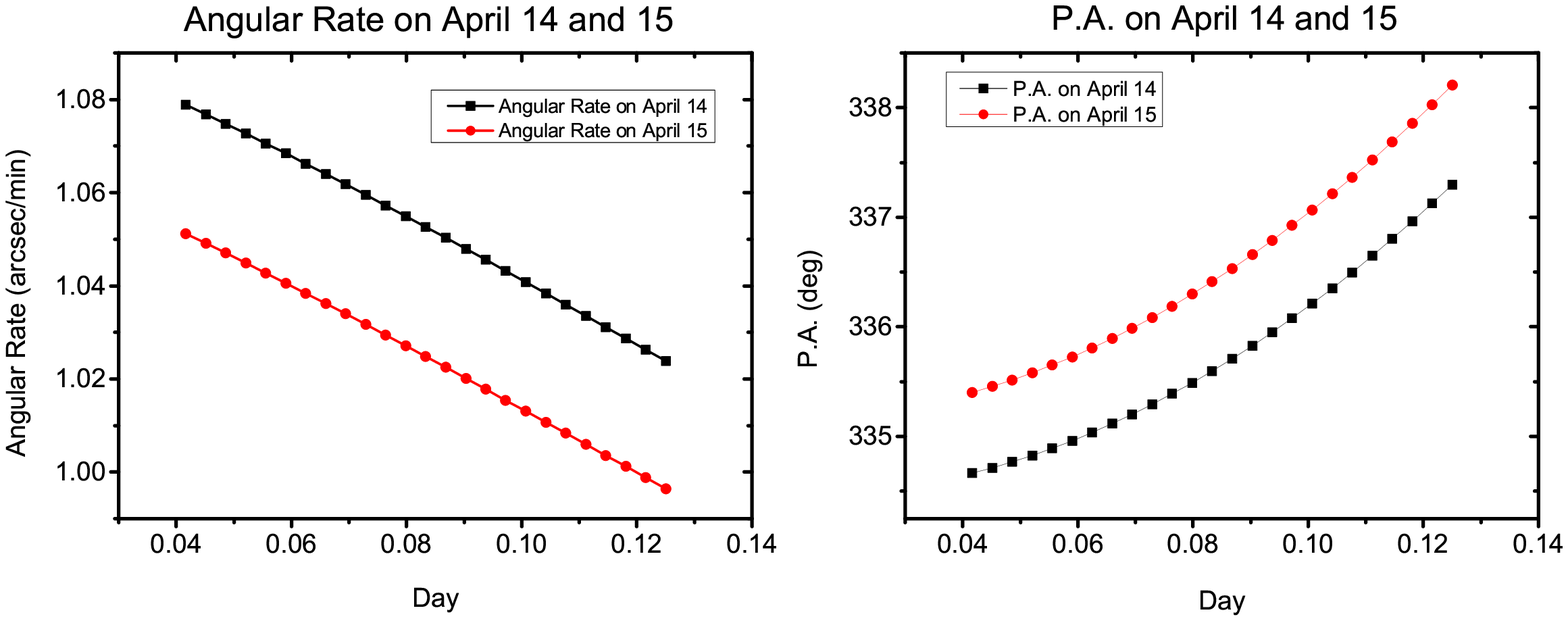}
	\caption{The angular rate (i.e. proper motions) and the position angle (P.A.) of Apophis according to JPL ephemeris (5 minutes for interval) during the observation time on April 14 and 15, 2021. The zero point of the time axis is set as UTC 12:00 on April 14, 2021. }
	\label{Fig5}
\end{figure}

\subsubsection{Stack of CCD Frames}
After alignment, the frames (including a master frame and nine individual frames in a group) are stacked with all the corresponding average pixel values using Drizzle method (\citealt{2002PASP..114..144F}). Then, we perform both star-stacking and object-stacking. In this way, we can derive the positions of stars from star-stacked frames and the positions of the object from object-stacked frames. The left and the right panel of Fig.\ref{Fig3} show a star-stacked frame and an object-stacked frame, respectively. We can see the trailing of Apophis in the star-stacked frame and the trailing of stars in the object-stacked frame. Fig.\ref{Fig4} shows the counts (ADU) across the pixel coordinate (in both X and Y axis) of Apophis in an individual frame and an object-stacked frame. In each panel, X Axis shows the pixel location of Apophis, and Y Axis shows the counts of the corresponding pixel locations. The sky backround is easier to be distinguished in the object-stacked frames. Compared with the panels in the individual frame, the distributions of the object-stacked frame are closer to Gaussian profile.

\subsubsection{Comparison with \textit{Astrometrica's} Stacking}

The stacking function of \textit{Astrometrica} is usually used to survey where we can set the angular rate (i.e. proper mostions) and the position angle (P.A.) of the moving object. However, the software assumes that the object goes along a uniform straight linear motion. Namely, the two parameters (angular rate and P.A.) have to be set as the constant in a group of frames. To obtain higher-precision postions of the object with a fast motion, we try to explore the effect of the changes of angular rate and P.A.. Fig.\ref{Fig5} shows the angular rate and the P.A. of Apophis according to JPL ephemeris changing over time during the observation. From Fig.\ref{Fig5}, the angular rate and P.A. change obviously with respect to the observation time. Compared with \textit{Astrometrica}, we calculate the relative shifts of each frame based on the object's ephemeris shifts and we consider the relative positions from ephemeris is accurate in a short time.

\begin{figure}
	\centering
	\includegraphics[width=.9\textwidth, angle=0]{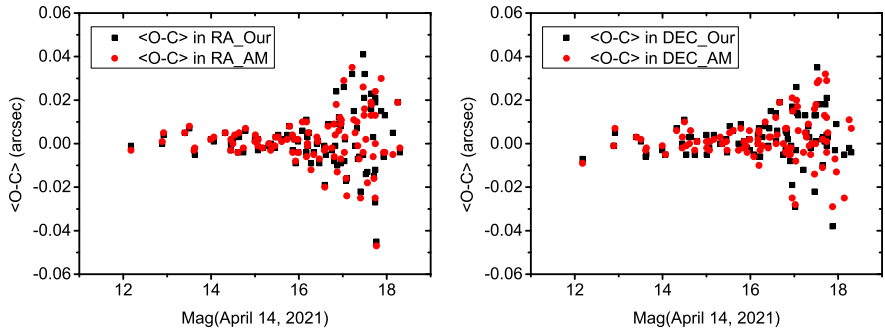}
	\caption{The left and the right panel show the common stars' mean (O-C)s of \textit{Astrometrica's} stacking and our stacking method, respectively based on the observations of April 14, 2021.}
	\label{Fig6}
\end{figure}

\begin{figure}
	\centering
	\includegraphics[width=.9\textwidth, angle=0]{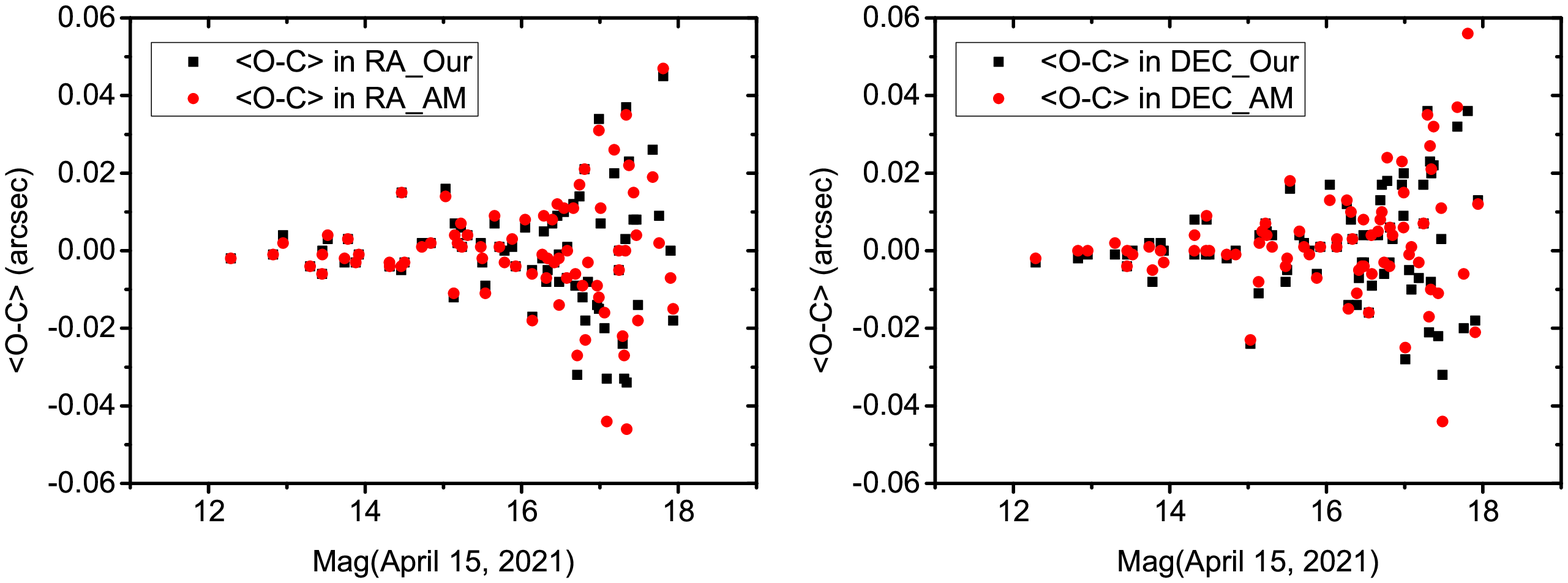}
	\caption{The left and the right panel show the common stars' mean (O-C)s of \textit{Astrometrica's} stacking and our stacking method, respectively based on the observations of April 15, 2021.}
	\label{Fig7}
\end{figure}

\begin{figure}
	\centering
	\includegraphics[width=.9\textwidth, angle=0]{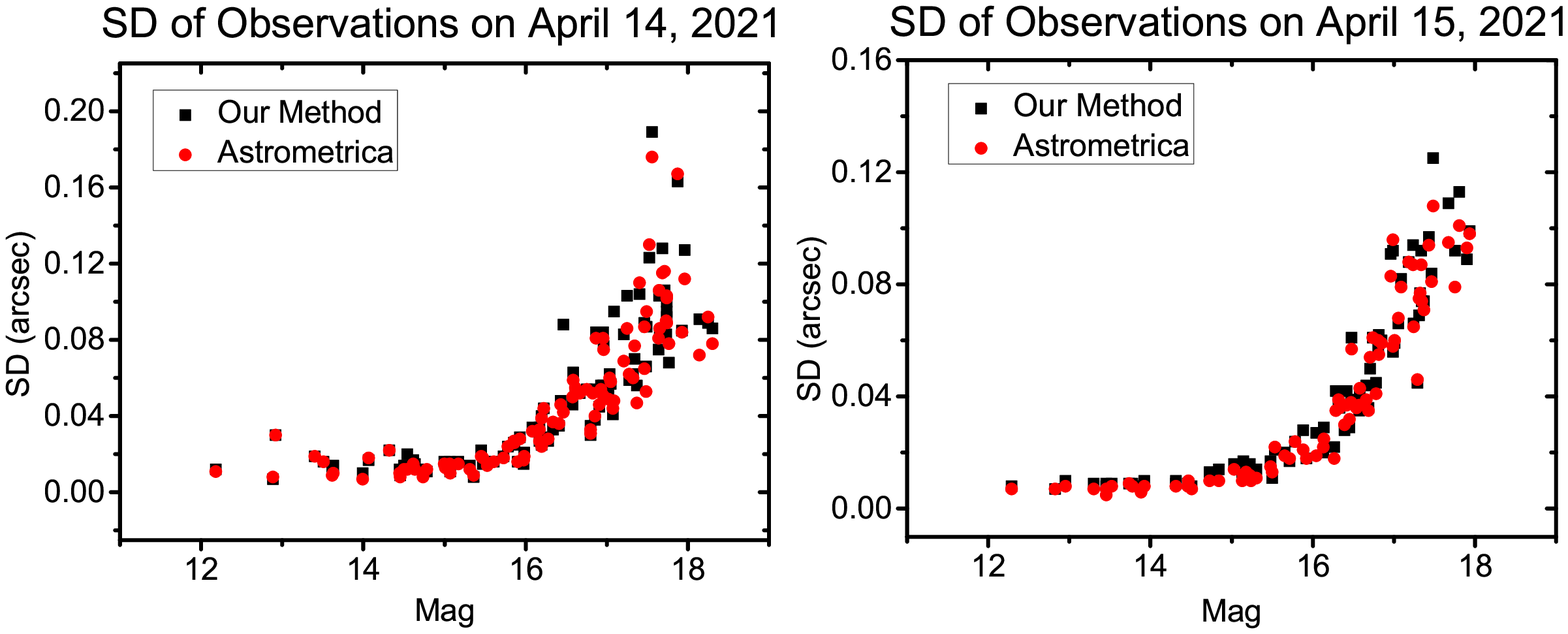}
	\caption{The left and the right panel show the common stars' SDs of \textit{Astrometrica's} stacking and our stacking method based on the observations of April 14, 2021 and April 15, 2021, respectively.}
	\label{Fig8}
\end{figure}

\begin{figure}
	\centering
	\includegraphics[width=.9\textwidth, angle=0]{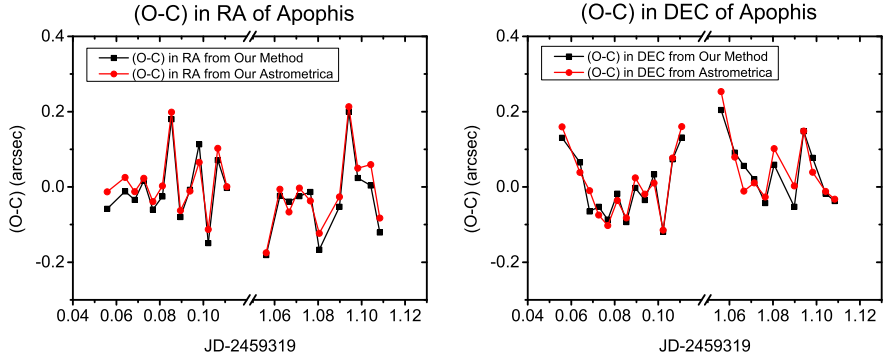}
	\caption{The left and the right panel show the object's (O-C)s of \textit{Astrometrica}'s stacking and our stacking method in right ascension and declination, respectively.}
	\label{Fig9}
\end{figure}

\subsubsection{Data Reduction}

We perform data reduction to the stacked observations from \textit{Astrometrica} and stacked observations based on the ephemeris shifts stacking, respectively. Each frame is reduced by the following procedures. Firstly, we measure the pixel positions $ (x, y) $ of each stars in the star-stacked frames by two-dimentional Gaussian centering algorithm. In the same way, we can obtain the pixel positions $ (x_{o}, y_{o}) $ of the object in the object-stacked frames. Secondly, we calculate the standard coordinate $ ( $$\xi$,$\eta$$ ) $ of each star through the central projection (\citealt{1985spas.book.....G}). The reference equatorial coordinates $ ( $$\alpha$,$\delta$$ ) $ here are taken from the newest Gaia EDR3 catalog (\citealt{2021A&A...649A...1G}) and calculated to the Astrometric positions at the observational epoch. The reference positions of the object (Apophis) are obtained from the JPL ephemeris. 101 and 82 Gaia stars are found for observations on April 14 and 15, respectively. The least squares scheme is used to solve the plate model with a weighted fourth-order polynomial (\citealt{2019MNRAS.490.4382L}). In this way, the (O-C)s (observed minus computed) of stars and the object can be calculated.

\subsubsection{Astrometric Results}

Fig.\ref{Fig6}-Fig.\ref{Fig8} show the mean (O-C)s and standard deviations (SD) of the stars of star-stacked frames in right ascension and declination stacked by \textit{Astrometrica} and stacked based on the ephemeris shifts, respectively. The two stacking methods show the consistency of both the accuracy and the precision. Fig.\ref{Fig9} shows the (O-C)s of the object (Apophis) by the two stacking methods in right ascension and declination, respectively and Table \ref{Tab:tab4} gives the mean (O-C)s and the SDs of the object. The observed topocentric astrometric positions of the object are listed in Table \ref{Tab:tab5}. Also, they can be downloaded from the website of Sino-French Joint Laboratory for Astrometry, Dynamics and Space Science of Jinan University\footnote{\it https://astrometry.jnu.edu.cn/download/list.htm}.

\begin{table}
	\begin{center}
		\caption[]{Statistics of mean (O-C)s and SDs for Apophis over Two Nights}\label{Tab:tab4}
		%%Please Capitalize the First Letter of Each Notional Word in table's caption
		\begin{tabular}{cccccccc}
			\hline\noalign{\smallskip}
			Date & Software & Mean (O-C) in RA (arcsec) & SD(arcsec) & Mean (O-C) in DEC (arcsec) & SD(arcsec) \\
			\hline\noalign{\smallskip}
			April 14, 2021 & Astrometrica  & 0.013   &  0.077  &   0.002  &  0.090          \\
			April 14, 2021 & This work           & -0.004   &  0.086  &  -0.003  &  0.084          \\
			April 15, 2021 & Astrometrica  & -0.017   &  0.103  &   0.050  &  0.089          \\
			April 15, 2021 & This work           & -0.036   &  0.104  &   0.046  &  0.082          \\
			Total          & Astrometrica  & -0.001   &  0.090  &   0.024  &  0.091          \\
			Total          & This work           & -0.018   &  0.094  &   0.020  &  0.085          \\
			
			\noalign{\smallskip}\hline
		\end{tabular}
	\end{center}
\end{table}

\begin{table}
	\begin{center}
		\caption[]{The observed topocentric astrometric positions of Apophis. The first column is the Julian Date and it is corresponding to the mid-time of each stacked frame. The observed topocentric astrometric positions in right ascention and declination of Apophis are listed in the second and the third column, respectively.}\label{Tab:tab5}
		%%Please Capitalize the First Letter of Each Notional Word in table's caption
		\begin{tabular}{ccc}
			\hline\noalign{\smallskip}
			JD & RA(h m s) & DEC($^{\circ}$  $'$  $''$) \\
			\hline\noalign{\smallskip}
			2459319.0557558 & 08   02   27.177   & +17   15   57.418              \\
			2459319.0640116 & 08   02   26.805   & +17   16    8.856             \\
			2459319.0684479 & 08   02   26.604   & +17   16   14.892              \\
			2459319.0726389 & 08   02   26.420   & +17   16   20.718             \\
			2459319.0768032 & 08   02   26.230   & +17   16   26.452              \\
			2459319.0811944 & 08   02   26.039   & +17   16   32.593             \\
			2459319.0853657 & 08   02   25.871   & +17   16   38.275             \\
			
			2459319.0894873  & 08   02   25.674    & +17   16   44.046           \\
			2459319.0938843  & 08   02   25.490    & +17   16   50.064          \\
			2459319.0980764  & 08   02   25.320    & +17   16   55.891         \\
			2459319.1022569  & 08   02   25.125    & +17   17    1.474         \\
			2459319.1066470  & 08   02   24.956    & +17   17    7.679        \\
			2459319.1108032  & 08   02   24.779    & +17   17   13.420        \\
			
			2459320.0561435  & 08   02    3.376    & +17   39   39.363        \\
			2459320.0624560  & 08   02    3.115    & +17   39   47.871        \\
			2459320.0666019  & 08   02    2.937    & +17   39   53.486        \\
			2459320.0713600  & 08   02    2.736    & +17   39   59.924        \\
			2459320.0764097  & 08   02    2.524    & +17   40    6.715        \\
			2459320.0805058  & 08   02    2.343    & +17   40   12.368        \\
			
			2459320.0897905  & 08   02    1.969    & +17   40   24.804        \\
			2459320.0940972  & 08   02    1.811    & +17   40   30.810       \\
			2459320.0982535  & 08   02    1.632    & +17   40   36.332       \\
			2459320.1042315  & 08   02    1.393    & +17   40   44.267       \\
			2459320.1083866  & 08   02    1.222    & +17   40   49.819       \\
			
			\noalign{\smallskip}\hline
		\end{tabular}
	\end{center}
\end{table}

From the results above, the dispersions of the two stacking methods show the consistency. \textit{Astrometrica} is an open well-integrated astrometric software and the difference of object's (O-C)s by the two methods might mainly derive from the considered velocity model. However, the difference of object's (O-C)s by two methods in such precision makes little sense to orbital determination. The results above also show that the object fainter than 18 mag can be detected through stacking. With high-credibility accuracy, the positions' precision of the object is better than 0.1 arcsec in both right ascension and declination.

\subsubsection{Discussion}
\label{sect:Discussion}
To detect unknown objects or those whose ephemeris is inaccurate, iterative stacking method (\citealt{2014ApJ...792...60Z}, \citealt{2015AJ....150..125H}) according to the stacked SNR of the object would be a better solution, which is promising to survey. However, this iterative method costs more time and computing resources of the computer. Besides, the astrometric precisions of the objects are to be explored because this method concentrates on detecting unknown object after all. \textit{Astrometrica} also provides the stacking function, assuming the object going along a uniform linear motion. The model of iterative stacking survey has not been supported so far. The iterative stacking method usually adopts the way of uniform straight linear motion for the objects. However, if the motion of the object changes significantly during the observation time such as NEAs, sometimes the effect of the changes of angular rate and its position angle should be considered. If the relative positions of the object from the ephemeris are accurate in a short time, we can take the effect of velocity change into account, which might help us obtain higher-precision astrometric positions of the object.
However, this work only tests the performance of the stacking object with angular rate of about 1 arcsec per minute. The performance of moving object with different angular rates hasn't been tested. The potential of this telescope at observing fast-moving objects hasn't been fully addressed.

\section{Conclusion and Outlook}
\label{sect:Conclusion}
In this paper, we have a test to the performance of the 80-cm azimuthal-mounting telescope at Yaoan Station, Purple Mountain Observatory in stability of GD, the limit magnitude and the astrometric accuracy and precision of the stack of fast-moving object Apophis. We find that the geometric distortion of the CCD is stable in a single epoch and multi epochs. From 8 derived GD solutions over about one year, the maximum values of each vector ranges from 0.75 to 0.79 pixel. The median value of each vector ranges from 0.14 to 0.16 pixel. About 20.5 magnitude (Gaia-G) stars can be detected with Johnson-V filter exposured in 300 seconds. The astrometric error is estimated at 0.14 arcsec using the fitted sigmoidal function. 24 groups of stacked observations are derived in total over two nights based on the ephemeris shifts of Apophis. After data reduction, our results show that the mean (O-C)s of Apophis are -0.018 and 0.020 arcsec in right ascention and declination, and the standard deviation are 0.094 and 0.085 arcsec, respectively. The results based on the ephemeris shifts have a little systematic deviation of object's (O-C)s compared with the results by uniform linear motion stacking of \textit{Astrometrica}, which is mainly probably caused by the velocity model of the object. The astrometric results show that the fast moving object fainter than 18 mag can be detected through stacking method with the precision better than 0.1 arcsec. However, this work only tests the stacking of object Apophis, and the potential of the telescope at observing fast-moving object hasn't been fully addressed. We are to explore the limit magnitude of the telescope with other filters (such as Johnson I filter) and objects with different angular rates for stacking. Of course, the telescope also has the potentiality to perform photometry, which will be explored with more proper observations.

\begin{acknowledgements}
We appreciate the reviewers (Val\'{e}ry Lainey and another anonymous reviewer) who provide the constructive comments. We are grateful for Drs. Li F., Yuan Y., and others for the group of 80-cm telescope providing the help of obtaining observations. We also thank Prof. Fu Y.N. from Purple Mountain Observatory and you give some valuable advice to our work. This research is supported by the National Natural Science Foundation of China (Grant No.11873026, 11273014), by the Joint Research Fund in Astronomy (Grant No.U1431227), by the science research grants from the China Manned Space Project with NO. CMS-CSST-2021-B08 and Excellent Postgraduate Recommendation Scientific Research Innovative Cultivation Program of Jinan University. This work has made use of data from the European Space Agency (ESA) mission \emph{Gaia} (\url{https://www.cosmos.esa.int/gaia}), processed by the \emph{Gaia} Data Processing and Analysis Consortium (DPAC, \url{https://www.cosmos.esa.int/web/gaia/dpac/consortium}). Funding for the DPAC has been provided by national institutions, in particular the institutions participating in the \emph{Gaia} Multilateral Agreement.
\end{acknowledgements}

\bibliographystyle{raa}
\bibliography{bibtex}

\begin{thebibliography}{24}
\providecommand\natexlab[1]{#1}
\providecommand\JournalTitle[1]{#1}

\bibitem[{Anderson} \& {King}(2003)]{2003PASP..115..113A}
{Anderson}, J., \& {King}, I.~R. 2003, \pasp, 115, 113

\bibitem[{Brozovi{\'c}} {et~al.}(2018)]{2018Icar..300..115B}
{Brozovi{\'c}}, M., {Benner}, L. A.~M., {McMichael}, J.~G., {et~al.} 2018,
  \icarus, 300, 115

\bibitem[{Farnocchia} {et~al.}(2013)]{2013Icar..224..192F}
{Farnocchia}, D., {Chesley}, S.~R., {Chodas}, P.~W., {et~al.} 2013, \icarus,
  224, 192

\bibitem[{French} {et~al.}(2006)]{2006PASP..118..246F}
{French}, R.~G., {McGhee}, C.~A., {Frey}, M., {et~al.} 2006, \pasp, 118, 246

\bibitem[{Fruchter} \& {Hook}(2002)]{2002PASP..114..144F}
{Fruchter}, A.~S., \& {Hook}, R.~N. 2002, \pasp, 114, 144

\bibitem[{Gaia~Collaboration} {et~al.}(2021)]{2021A&A...649A...1G}
{Gaia~Collaboration}, {Brown}, A.~G.~A., {Vallenari}, A., {et~al.} 2021, \aap,
  649, A1

\bibitem[{Giorgini} {et~al.}(2008)]{2008Icar..193....1G}
{Giorgini}, J.~D., {Benner}, L. A.~M., {Ostro}, S.~J., {Nolan}, M.~C., \&
  {Busch}, M.~W. 2008, \icarus, 193, 1

\bibitem[{Green}(1985)]{1985spas.book.....G}
{Green}, R.~M. 1985, {Spherical Astronomy}

\bibitem[{Heinze} {et~al.}(2015)]{2015AJ....150..125H}
{Heinze}, A.~N., {Metchev}, S., \& {Trollo}, J. 2015, \aj, 150, 125

\bibitem[{Li} \& {Peng}(2020)]{2020AcASn..61...28L}
{Li}, C.~W., \& {Peng}, Q.~Y. 2020, Acta Astronomica Sinica, 61, 28

\bibitem[{Li} {et~al.}(2021)]{2021SSPMA..51b9502L}
{Li}, F., {Yuan}, Y., \& {Fu}, Y. 2021, Scientia Sinica Physica, Mechanica \&
  Astronomica, 51, 029502

\bibitem[{Lin} {et~al.}(2019)]{2019MNRAS.490.4382L}
{Lin}, F.~R., {Peng}, J.~H., {Zheng}, Z.~J., \& {Peng}, Q.~Y. 2019, \mnras,
  490, 4382

\bibitem[{Lindegren}(1980)]{1980A&A....89...41L}
{Lindegren}, L. 1980, \aap, 89, 41

\bibitem[{Peng} {et~al.}(2012)]{2012AJ....144..170P}
{Peng}, Q.~Y., {Vienne}, A., {Zhang}, Q.~F., {et~al.} 2012, \aj, 144, 170

\bibitem[{Shao} {et~al.}(2014)]{2014ApJ...782....1S}
{Shao}, M., {Nemati}, B., {Zhai}, C., {et~al.} 2014, \apj, 782, 1

\bibitem[{Souchay} {et~al.}(2018)]{2018A&A...617A..74S}
{Souchay}, J., {Lhotka}, C., {Heron}, G., {et~al.} 2018, \aap, 617, A74

\bibitem[{Thuillot} {et~al.}(2015)]{2015A&A...583A..59T}
{Thuillot}, W., {Bancelin}, D., {Ivantsov}, A., {et~al.} 2015, \aap, 583, A59

\bibitem[{Tyson} {et~al.}(1992)]{1992AAS...181.0610T}
{Tyson}, J.~A., {Guhathakurta}, P., {Bernstein}, G.~M., \& {Hut}, P. 1992, in
  American Astronomical Society Meeting Abstracts, Vol. 181, American
  Astronomical Society Meeting Abstracts, 06.10

\bibitem[{Vokrouhlick{\'y}} {et~al.}(2015)]{2015Icar..252..277V}
{Vokrouhlick{\'y}}, D., {Farnocchia}, D., {{\v{C}}apek}, D., {et~al.} 2015,
  \icarus, 252, 277

\bibitem[{Wang} {et~al.}(2017)]{2017AcASn..58...49W}
{Wang}, B., {Zhao}, H.~B., \& {Li}, B. 2017, Acta Astronomica Sinica, 58, 49

\bibitem[{Wang} {et~al.}(2015)]{2015MNRAS.454.3805W}
{Wang}, N., {Peng}, Q.~Y., {Zhang}, X.~L., {et~al.} 2015, \mnras, 454, 3805

\bibitem[{Yelda} {et~al.}(2010)]{2010ApJ...725..331Y}
{Yelda}, S., {Lu}, J.~R., {Ghez}, A.~M., {et~al.} 2010, \apj, 725, 331

\bibitem[{Yuan} {et~al.}(2021)]{2021A&A...645A..48Y}
{Yuan}, Y., {Li}, F., {Fu}, Y., \& {Ren}, S. 2021, \aap, 645, A48

\bibitem[{Zhai} {et~al.}(2014)]{2014ApJ...792...60Z}
{Zhai}, C., {Shao}, M., {Nemati}, B., {et~al.} 2014, \apj, 792, 60

\end{thebibliography}

\label{lastpage}

\end{document}